\pdfoutput=1
\documentclass[11pt]{article}
\usepackage[utf8]{inputenc}
\usepackage[top=70pt,bottom=70pt,left=60pt,right=60pt]{geometry}
\usepackage{amsmath}
\usepackage{hyperref}
\usepackage{amssymb}
\usepackage{graphicx,caption,subcaption}
\usepackage{cite}
\usepackage{bm}
\usepackage{ mathrsfs }
\usepackage[space]{grffile}
\usepackage{graphicx}
\newcommand{\overbar}[1]{\mkern 1.5mu\overline{\mkern-1.5mu#1\mkern-1.5mu}\mkern 1.5mu}

\begin{document}
\thispagestyle{empty}

{\hbox to\hsize{
\vbox{\noindent December 2019 (revised)}}}

\noindent
\vskip2.0cm
\begin{center}

{\Large\bf Aspects of gauged R-symmetry in $SU(1,1)/U(1)$ supergravity}

\vglue.3in

Yermek Aldabergenov~${}^{a,b}$
\vglue.1in

${}^a$~Department of Physics, Faculty of Science, Chulalongkorn University,\\
Thanon Phayathai, Pathumwan, Bangkok 10330, Thailand\\
${}^b$~Institute of Experimental and Theoretical Physics, Al-Farabi Kazakh National University, \\
71 Al-Farabi Avenue, Almaty 050040, Kazakhstan \\
\vglue.1in
yermek.a@chula.ac.th
\end{center}

\vglue.3in

\begin{center}
{\Large\bf Abstract}
\vglue.2in
\end{center}

We propose a novel realization of spontaneous supersymmetry breaking in de Sitter vacuum by F- and D-terms in $N=1$ four-dimensional supergravity coupled to a chiral superfield with $SU(1,1)/U(1)$ target space. Our construction features gauged $U(1)_R$ symmetry rotating the chiral scalar field by a phase. Both SUSY and R-symmetry can be spontaneously broken, and for two particular parameter choices we obtain no-scale supergravity with positive tunable cosmological constant.

\newpage

\section*{Introduction}

Supersymmetry (SUSY) is a compelling idea that is motivated by both phenomenological (Beyond the Standard Model) and theoretical (String Theory) point of view. If nature indeed uses supersymmetry it must be spontaneously broken. In the simplest scenario SUSY breaking happens in the hidden sector and is mediated to the visible sector (Supersymmetric Standard Model) by gravitational interactions. It is therefore of interest to study SUSY breaking in the context of $N=1$ four-dimensional supergravity (SUGRA).

On the other hand, according to observations the Universe is currently expanding with acceleration \cite{Riess:1998cb,Perlmutter:1997zf}. The simplest way to describe such a universe is by introducing a (very) small positive cosmological constant. In supergravity the task of adding a positive cosmological constant is known to be non-trivial because of the restrictions on the scalar potential imposed by supersymmetry. For example in pure (standard) supergravity one can only have zero (Minkowski vacuum) or negative (anti-de Sitter vacuum) cosmological constant \cite{Townsend:1977qa}. It is possible to generate a positive cosmological constant if we allow other (non-gravitational) multiplets. One interesting possibility is that the same field(s) that breaks SUSY can also generate the cosmological constant. This is possible, for example, in the simplest Polonyi model \cite{Polonyi:1977pj,Linde:2016bcz,Aldabergenov:2017bjt}.

In this work we will focus on the supergravity non-linear $\sigma$-model with $SU(1,1)/U(1)$ target space. This coset manifold, known as the Poincar\'e plane, describes hyperbolic K\"ahler geometry, and often arises in superstring-derived effective SUGRA models where the corresponding scalars are the compactification moduli. Our goal is to find a Poincar\'e plane model that spontaneously breaks supersymmetry in de Sitter vacuum, i.e. allowing for a positive (tunable) cosmological constant. It turns out, one such class of models is available if we introduce linearly realized gauged $U(1)_R$ symmetry. This, of course, adds a gauge (vector) multiplet with its $D$-term contribution to the scalar potential and SUSY breaking.

This paper is organized as follows. In Section 1 we recall basic properties of $N=1$ four-dimensional supergravity as well as the $SU(1,1)/U(1)$ non-linear $\sigma$-model. We discuss the two equivalent coordinate choices -- one covering the whole Poincar\'e plane (disk) while the other covering its upper half. In Section 2 we use the fact that the two parametrizations of the plane reveal two different types of $U(1)$ symmetries (linearly and non-linearly realized), to construct new models where the $U(1)$ is linearly realized local R-symmetry. In Section 3 we show that for suitable parameter choices our models spontaneously break SUSY and R-symmetry, and generate tunable cosmological constant. We find that in two particular cases the scalar potential becomes flat with positive height (de Sitter no-scale supergravity). Some generalizations of the our models are discussed in Section 4, while Section 5 is devoted for further discussion and conclusion.

\section{\texorpdfstring{$N=1$ $D=4$}{Lg} supergravity and the Poincar\'e plane}

Let us briefly review the general features of the standard four-dimensional $N=1$ supergravity. Its bosonic sector is described by the action (we use Planck units, $\kappa=1$, unless otherwise stated)~\footnote{A derivation of this action can be found in Refs. \cite{Wess:1992cp,Freedman:2012zz}}
\begin{equation}
    e^{-1}{\cal L}=\frac{1}{2}R-K_{i\bar{j}}D_m\Phi^i\overbar{D^m\Phi}^j-\frac{1}{4}f^R_{AB}F_{mn}^AF^{B,mn}-\frac{i}{4}f^I_{AB}\tilde{F}_{mn}^AF^{B,mn}-V_F-V_D~,\label{standardaction}
\end{equation}
whose the F- and D- type scalar potentials are given by
\begin{gather}
    V_F=e^K\left[K^{i\bar{j}}(W_i+K_iW)(\overbar{W}_{\bar{j}}+K_{\bar{j}}\overbar{W})-3|W|^2\right]~,\label{VF}\\
    V_D=\frac{g^2}{2}f_R^{AB}\mathscr{D}_A\mathscr{D}_B~,\label{VD}
\end{gather}
where $K=K(\Phi_i,\overbar{\Phi}_i)$ is a (real) K\"ahler potential depending upon chiral scalar fields $\Phi_i$, $W=W(\Phi_i)$ is a (holomorphic) superpotential, $f_{AB}=f_{AB}(\Phi_i)$ is a (holomorphic) gauge kinetic function with $f^R_{AB}\equiv {\rm Re}f_{AB}$ and $f^I_{AB}\equiv {\rm Im}f_{AB}$; $R$ is the spacetime scalar curvature, $F_{mn}^A=\partial_mA_n^A-\partial_nA_m^A+gf^{ABC}A^B_mA^C_n$ is the field strength of a vector (gauge) field $A_m^A$, $g$ is the gauge coupling, and $\mathscr{D}_A$ are Killing potentials of the gauged isometries of the K\"ahler manifold. We use the notation $K^{i\bar{j}}\equiv K_{i\bar{j}}^{-1}$, where $K_{i\bar{j}}\equiv\frac{\partial^2K}{\partial\Phi_i\partial\overbar{\Phi}_j}$, $W_i\equiv\frac{\partial W}{\partial\Phi_i}$, and $f^{AB}\equiv f_{AB}^{-1}$ with $A,B$ as the gauge group indices. The gauge-covariant derivatives of the charged scalars are
\begin{equation}
    D_m\Phi^i=\partial_m\Phi^i-gA_m^AX_A^i~,\label{DPhi}
\end{equation}
where $X^i_A$ are the corresponding Killing vectors.

The action \eqref{standardaction} is invariant under combined K\"ahler-Weyl transformations
\begin{equation}
    K\rightarrow K+\Sigma+\overbar{\Sigma}~,~~~W\rightarrow We^{-\Sigma}~,\label{KWtransform}
\end{equation}
where $\Sigma$ is an arbitrary chiral scalar field.

Killing potentials can be related to Killing vectors by the expression
\begin{equation}
    \mathscr{D}_A=i\left(K_i+\frac{W_i}{W}\right)X^i_A~,\label{Killingpot}
\end{equation}
where the superpotential-dependent term is present whenever R-symmetry is gauged, and is known as the Fayet-Iliopoulos term (of gauged R-symmetry) in supergravity.

SUSY is spontaneously broken whenever auxiliary $F$ and/or $D$ fields, satisfying
\begin{gather}
    F^i=-e^{K/2}K^{i\bar{j}}(\overbar{W}_{\bar{j}}+K_{\bar{j}}\overbar{W})~,\label{F_aux}\\
    D_A=-g\mathscr{D}_A~,\label{D_aux}
\end{gather}
acquire non-vanishing VEVs. When SUSY is broken gravitino becomes massive absorbing the goldstino. In the Lagrangian the gravitino effective mass appears as
\begin{equation}
    m_{3/2}^2=e^K|W|^2~.\label{m_32}
\end{equation}
In Minkowski background the VEV of $m_{3/2}$ is the physical gravitino mass, however in more complicated backgrounds physical mass differs from the "Lagrangian" mass given by Eq. \eqref{m_32}. Throughout the paper we will use the term "gravitino mass" in the sense of Eq. \eqref{m_32}.~\footnote{One can borrow the notion of the physical gravitino mass from AdS supergravity as $m_{3/2,{\rm phys}}^2=\langle m_{3/2}\rangle^2 +V_0/3$ (see e.g. \cite{Freedman:2012zz} and Refs. therein). In (pure) AdS supergravity the cosmological constant is $V_0=-3\langle m^2_{3/2}\rangle$ and the physical mass vanishes.} Then, $\langle m_{3/2}\rangle$ can be zero even when SUSY is broken.

As regards the Poincar\'e plane, it can be described by the K\"ahler metric in terms of the half-plane coordinate $T$ (a complex scalar in spacetime) as
\begin{equation}
    K_{T\overbar{T}}=\frac{\alpha}{(T+\overbar{T})^2}~,\label{metricT}
\end{equation}
with some positive real number $\alpha$ that determines the K\"ahler curvature, $R_K=-2/\alpha$. Alternatively, the same metric can be defined using the disk coordinate $Z$,
\begin{equation}
    K_{Z\overbar{Z}}=\frac{\alpha}{(1-Z\overbar{Z})^2}~.\label{metricZ}
\end{equation}
The two metrics are related by the Cayley transformation 
\begin{equation}
    Z=\frac{T-1}{T+1}~.\label{ZT}
\end{equation}

From string theory point-of-view, the Poincar\'e plane models corresponding to compactification moduli have (positive) integer values of $\alpha$. In principle, the available values are $\alpha=1,2,...,7$ according to Refs. \cite{Duff:2010ss,Duff:2010vy,Ferrara:2016fwe}.

The metric \eqref{metricZ} can be obtained from the K\"ahler potential $K=-\alpha\log(1-Z\overbar{Z})$. Under the transformation \eqref{ZT} it becomes
\begin{equation}
    K=-\alpha\left[\log(T+\overbar{T})-\log(T+1)-\log(\overbar{T}+1)\right]~,
\end{equation}
plus an irrelevant constant. The last two terms can be absorbed into the superpotential by the K\"ahler-Weyl transformation \eqref{KWtransform} with $\Sigma=-\alpha\log(T+1)$. To summarize, assuming the general superpotential $W=W(Z)$, the transformation \eqref{ZT} followed by the K\"ahler-Weyl rescaling takes the $Z$-parametrization of the Poincar\'e plane to the (equivalent) $T$-parametrization as follows
\begin{equation}
    \begin{cases}
    K=-\alpha\log(1-Z\overbar{Z})\\
    W=W(Z)
    \end{cases}~\Longrightarrow~\begin{cases}
    K=-\alpha\log(T+\overbar{T})\\
    W=W\left(\frac{T-1}{T+1}\right)(T+1)^\alpha~.
    \end{cases}\label{ZTKahler}
\end{equation}

The Poincar\'e plane has a wide range of applications in phenomenology. For example, the choice $K=-3\log(T+\overbar{T})$ and $W=W_0$ ($W_0$ is a constant) corresponds to the simplest no-scale supergravity \cite{Cremmer:1983bf,Ellis:1983sf,Ellis:1983ei}. Using the inverse transformation of Eq. \eqref{ZT} the no-scale model can be expressed in terms of the disk coordinate $Z$ as $K=-3\log(1-Z\overbar{Z})$ and $W=W_0 (Z-1)^3$. 

In the both coordinate choices ($T$ and $Z$) the complex scalars can be parametrized in such a way that one of their two real components is canonical. $T$ can be parametrized as
\begin{equation}
    T=\frac{1}{2}e^{-\sqrt{\frac{2}{\alpha}}\varphi}+it~,\label{Tpar}
\end{equation}
where the real scalar $\varphi$ is canonical, while $t$ (also real) is not -- its kinetic term is coupled to $\varphi$. The disk coordinate $Z$ can be parametrized e.g. in a polar form,
\begin{equation}
    Z=e^{-i\zeta}\tanh{\frac{\phi}{\sqrt{2\alpha}}}~,\label{Zpar}
\end{equation}
where $\phi$ is the canonical scalar controlling the absolute value of $Z$, and $\zeta$ is the scalar controlling its angle. This parametrization of $Z$ will be useful in the following sections.

\section{Gauged R-symmetry in \texorpdfstring{$SU(1,1)/U(1)$}{Lg} models}

$U(1)$ gauge theories in the context of $SU(1,1)/U(1)$ models are often considered as half-plane models with the K\"ahler potential
\begin{equation}
    K=-\alpha\log(T+\overbar{T})~,
\end{equation}
where the symmetry under imaginary shifts of $T$ is gauged. The local shifts can be written as $T\rightarrow T+iq_T\theta$, where $\theta=\theta(x)$ is the gauge parameter and $q_T$ is the corresponding $U(1)$ charge of $T$. The Killing vector must satisfy the relation $\delta T=\theta X^T$, thus $X^T=iq_T$.

If we want to promote this gauge transformation to a local R-transformation, superpotential must transform as
\begin{equation}
    W\rightarrow We^{-iq\theta}~,\label{W(T)transform}
\end{equation}
where $q$ is the $U(1)_R$ charge of the superpotential. If there are no other chiral fields in the model, the superpotential is fixed as $W=\mu e^{-\xi T}$ with some real constant $\xi$ and complex constant $\mu$. From the transformation property \eqref{W(T)transform} we obtain the relation $\xi=q/q_T$. Eq. \eqref{Killingpot} in this case yields
\begin{equation}
    {\mathscr D}=q_T\left(\frac{\alpha}{T+\overbar{T}}+\xi\right)~,
\end{equation}
which makes it clear that $\xi$ is exactly the FI term of gauged R-symmetry that we mentioned earlier.

If we switch to the $Z$-parametrization of the Poincar\'e plane with
\begin{equation}
    K=-\alpha\log(1-Z\overbar{Z})~,
\end{equation}
the phase symmetry of $Z$ becomes the simplest choice for gauging. I.e. we can introduce the gauge transformation $Z\rightarrow Ze^{-iq_Z\theta}$, where $q_Z$ is the $U(1)$ charge of $Z$, with the corresponding Killing vector $X^Z=-iq_ZZ$. Promoting this transformation to an R-transformation, as usual, requires that the superpotential transforms as in Eq. \eqref{W(T)transform}. This fixes the superpotential as $W=\mu Z^n$ where $n=q/q_Z$. To avoid negative powers of $Z$ in the action $n$ must be greater or equal to one (unlike negative powers of $T$ in the half-plane case, negative powers of $Z$ lead to singularities as can be seen from parametrizations \eqref{Tpar} and \eqref{Zpar}). The Killing potential now takes the form
\begin{equation}
    {\mathscr D}=q_Z\left(\frac{\alpha Z\overbar{Z}}{1-Z\overbar{Z}}+n\right)~,\label{KillingZ}
\end{equation}
with $n$ as the FI term. Let us investigate this setup in more detail.

\section{Properties of the scalar potential}

Our model of interest is defined by~\footnote{Similar setup was considered in Ref. \cite{Pallis:2018xmt} in the context of SUSY breaking, but without gauging the R-symmetry.}
\begin{gather}
    K=-\alpha\log(1-Z\overbar{Z})~,\\
    W=\mu Z^n~.\label{WZn}
\end{gather}
The superpotential is fixed by requiring R-symmetry, and for simplicity we put $n=1$ and $q=q_Z=1$ (the notation is the same as in the previous section). Also, without loss of generality we can consider $\mu$ to be real. Upon gauging the R-symmetry the Killing potential \eqref{KillingZ} is generated. After choosing the simplest gauge kinetic function $f=1$, we calculate the full scalar potential $V=V_F+V_D$,
\begin{gather}
    V_F=\mu^2\frac{(\alpha-1)^2z^4-(\alpha+2)z^2+1}{\alpha(1-z^2)^\alpha}~,\\
    V_D=\frac{g^2}{2}\left(\frac{\alpha z^2}{1-z^2}+1\right)^2~,
\end{gather}
where for convenience we introduced the notation $z\equiv |Z|$. When using the parametrization \eqref{Zpar} the angular mode $\zeta$ conveniently drops out of the scalar potential, and $z=\tanh\frac{\phi}{\sqrt{2\alpha}}$.

We can find critical points of the potential by studying the equation
\begin{equation}
    \frac{dV}{dz}=2z\frac{\left[(\alpha-1)z^2+1\right]\left[\alpha^2g^2(1-z^2)^\alpha+\mu^2(1-z^2)^2\left((\alpha-2)(\alpha-1)z^2-2\right)\right]}{\alpha(1-z^2)^{\alpha+3}}=0~.\label{dVdz}
\end{equation}
Regardless of the value of $\alpha$ there is always a critical point at $z=0$, where the scalar potential reduces to
\begin{equation}
    V(z=0)=\frac{\mu^2}{\alpha}+\frac{g^2}{2}~.\label{V_z=0}
\end{equation}

The equation for critical points other than $z=0$ can be reduced from Eq. \eqref{dVdz} to
\begin{equation}
    \alpha^2g^2(1-z^2)^\alpha+\mu^2(1-z^2)^2\left((\alpha-2)(\alpha-1)z^2-2\right)=0~,\label{zcrit}
\end{equation}
because the expression in the first square brackets of \eqref{dVdz} is non-vanishing even when $\alpha<1$, thanks to the canonical normalization $z^2=\tanh^2(\phi/\sqrt{2\alpha})<1$.

The existence of consistent solutions to Eq. \eqref{zcrit} depends on the choice of $\alpha$. First, let us consider the cases $\alpha=1,2,3,4$, as they can be studied analytically (we will comment on more general $\alpha$ in the next section).

$\bm{\alpha=1}$. Here the solution for Eq. \eqref{zcrit} is $z^2=1-\frac{g^2}{2\mu^2}$. This solution is valid if $2\mu^2>g^2$ in which case it corresponds to two minima (with $Z_2$ symmetry) while $z=0$ is a local maximum. Then the R-symmetry is spontaneously broken due to non-vanishing superpotential, while SUSY is broken due to~\footnote{For convenience we dropped the minus signs on the RHS in Eqs. \eqref{F_aux} and \eqref{D_aux}.}
\begin{gather}
    \langle F\rangle=g/\sqrt{2}~,~~~\langle D\rangle=2\mu^2/g~,\label{FD_alpha=1}\\
    \langle m_{3/2}\rangle^2=\frac{2\mu^4}{g^2}\left(1-\frac{g^2}{2\mu^2}\right)~,
\end{gather}
and the following cosmological constant is generated,
\begin{equation}
    V_0=\frac{\mu^2}{g^2}(3g^2-2\mu^2)~,\label{CC_alpha=1}
\end{equation}
so that we have AdS minimum if $3g^2<2\mu^2$, Minkowski minimum if $3g^2=2\mu^2$, and dS minimum if $6\mu^2>3g^2>2\mu^2$ (the first inequality ensures $z^2>0$). These conditions show that if we want Minkowski or de Sitter vacuum, both $F$- and $D$-term contributions \eqref{FD_alpha=1} to SUSY breaking must be comparable in magnitude. As $U(1)_R$ is spontaneously broken, the Killing vector $X^Z=-iZ$ is non-vanishing at the minimum. This generates a mass term for the gauge boson proportional to $g^2\langle Z\rangle^2$, as can be seen from Eq. \eqref{DPhi}, while the goldstone mode $\zeta$ can be gauged away. As for the mass of the canonical scalar $\phi$, after introducing its excitation $\delta\phi\equiv\phi-\phi_0$ and expanding the potential around the minimum, it reads
\begin{equation}
    m^2_{\delta\phi}=\frac{8\mu^4}{g^2}\left(1-\frac{g^2}{2\mu^2}\right)~,\label{m_deltaphi}
\end{equation}
which is positive since $2\mu^2>g^2$, and is twice the gravitino mass, $m_{\delta\phi}=2\langle m_{3/2}\rangle$.

In order to describe dark energy, $V_0$ must be positive and very small, namely $V_0\sim 10^{-120}$ in Planck units. From Eq. \eqref{CC_alpha=1} it is clear that this can be achieved in two ways. The first option is to set $\mu^2\sim 10^{-120}$, which will also force $g^2\sim 10^{-120}$ as required by the dS condition $6\mu^2>3g^2>2\mu^2$. This is phenomenologically problematic, as it means that SUSY breaking scale is of the same order as the dark energy scale. A more viable option is the fine tuning of the difference $3g^2-2\mu^2$ so that it almost vanishes. This does not require the individual parameters $g$ and $\mu$ -- and thus the SUSY breaking scale -- to be small. The relation $3g^2\approx 2\mu^2$ then simplifies the gravitino and scalar masses as
\begin{equation}
    \langle m_{3/2}\rangle^2\approx 3g^2~,~~~m_{\delta\phi}^2\approx 12g^2~.
\end{equation}

When $2\mu^2\leq g^2$ the solution $z^2=1-g^2/(2\mu^2)$ does not exist and the point $z=0$ is the global minimum (with no other critical points). In such case SUSY is broken by $\langle F\rangle=\mu$ and $\langle D\rangle=g$ while R-symmetry is restored at the minimum since the superpotential vanishes. This means that the gravitino mass $\langle m_{3/2}\rangle$, as well as the masses of the $U(1)_R$ gauge boson and the $\zeta$ scalar, are zero. This scenario is not viable from phenomenological point of view because there is a massless scalar in the spectrum, and the scales of SUSY breaking and the cosmological constant are identified.

$\bm{\alpha=2}$. In this case $z=0$ is the only critical point: if $2g^2>\mu^2$ it is a de Sitter minimum (with broken SUSY and unbroken R-symmetry), if $2g^2<\mu^2$ it is a maximum and the potential is unbounded from below. When $2g^2=\mu^2$, however, the potential is flat -- we have a no-scale model in de Sitter spacetime with the cosmological constant $V=3g^2/2$. The VEVs of $F$- and $D$-terms are
\begin{equation}
    \langle F\rangle=\frac{g}{\sqrt{2}}(1+z_0^2)~,~~~\langle D\rangle=g\frac{1+z_0^2}{1-z_0^2}~,
\end{equation}
where $z_0$ (the VEV of $z$) is arbitrary at the classical level. Thus, SUSY and R-symmetry are broken (as long as $z_0\neq 0$). The fact that $z^2=\tanh^2{(\phi/\sqrt{2\alpha})}$ has the range $0\leq z^2<1$ implies that
\begin{align}
    \frac{g}{\sqrt{2}}\leq \langle F\rangle &<\sqrt{2}g~,\\
    g\leq \langle D\rangle &<\infty~.
\end{align}
Small cosmological constant requires proportionally small $g^2$. Then $\langle F\rangle$ must also be small because it is proportional to $g$, but $\langle D\rangle$ can take large values if $z_0^2$ is close to one. The same is true for the gravitino mass,
\begin{equation}
    \langle m_{3/2}\rangle^2=\frac{2g^2z_0^2}{(1-z_0^2)^2}~.
\end{equation}

$\bm{\alpha=3}$. Similarly to the $\alpha=2$ case, when $\alpha=3$ there is only one critical point, $z=0$, and if $9g^2>2\mu^2$ it is a dS minimum, whereas if $9g^2<2\mu^2$ it is a maximum. If $9g^2=2\mu^2$ we once again arrive at a no-scale de Sitter model, this time with the cosmological constant $V=2g^2$. The auxiliary fields and the gravitino mass at the minimum are
\begin{gather}
    \langle F\rangle=\frac{g}{\sqrt{2}}\frac{1+2z_0^2}{\sqrt{1-z^2_0}}~,~~~\langle D\rangle=g\frac{1+2z^2_0}{1-z_0^2}~,\\
    \langle m_{3/2}\rangle^2=\frac{9g^2z_0^2}{2(1-z_0^2)^3}~.
\end{gather}
and have the following range
\begin{align}
    \frac{g}{\sqrt{2}}\leq \langle F\rangle &<\infty~,\\
    g\leq \langle D\rangle &<\infty~,
\end{align}
while $\langle m_{3/2}\rangle$ can take any value from zero (when $z_0=0$) to infinity (when $|z_0|\rightarrow 1$). Unlike the previous case, here both $\langle F\rangle$ and $\langle D\rangle$ can be large regardless of the value of $g$, if $z_0$ is close to one. However, in both $\alpha=2$ and $\alpha=3$ cases the $D$-term VEV necessarily dominates, $\langle D\rangle\gtrsim\langle F\rangle$.

$\bm{\alpha=4}$. In this case Eq. \eqref{zcrit} is solved by
\begin{equation}
    z^2=\frac{1}{2A}\left(2A-3+\sqrt{9-8A}\right)~,~~~A\equiv\frac{8g^2}{\mu^2}~.\label{z0_and_A}
\end{equation}
This is complemented by the condition
\begin{equation}
    0<A<1~\Longrightarrow~0<g^2<\mu^2/8~,\label{A_condition}
\end{equation}
that ensures that $z^2>0$. The cosmological constant corresponding to this minimum reads
\begin{equation}
    V_0=\frac{g^2}{2\mu^2}(9\mu^2-32g^2)~.
\end{equation}
If we require $V_0$ to be very small, the only choice is $g\ll 1$, because the cancellation $9\mu^2-32g^2\approx 0$ is incompatible with the condition \eqref{A_condition}.

F-/D-terms and the gravitino mass are non-vanishing,
\begin{gather}
    \langle F\rangle=\frac{\mu}{4}\sqrt{9-8A}~,~~~\langle D\rangle=g\sqrt{9-8A}~,\\
    \langle m_{3/2}\rangle^2=8\mu^2A^3\frac{2A-3+\sqrt{9-8A}}{(-3+\sqrt{9-8A})^4}~.\label{m32_alpha=4}
\end{gather}
Since $A$ ranges from zero to one, we have
\begin{align}
    \frac{\mu}{4}<\langle F\rangle &<\frac{3\mu}{4}~,\\
    g<\langle D\rangle &<3g~.
\end{align}
Also $\langle F\rangle>\langle D\rangle/\sqrt{2}$, due to the condition \eqref{A_condition}. If  $g\ll 1$, as required to describe dark energy, $\langle D\rangle$ becomes small, but there is still a freedom to control $\langle F\rangle$ and $\langle m_{3/2}\rangle$ by choosing the parameter $\mu$. In particular, the gravitino mass \eqref{m32_alpha=4} can be expanded in the limit $g\rightarrow 0$ (or $A\rightarrow 0$) as
\begin{equation}
    \langle m_{3/2}\rangle^2\approx \frac{27}{16}\mu^2~.
\end{equation}

As regards the scalar mass, it reads
\begin{equation}
    m^2_{\delta\phi}=\frac{\mu^2}{32}(9-8A)(3-4A+\sqrt{9-8A})~,\label{m_deltaphi_2}
\end{equation}
where $A$ is defined in Eq. \eqref{z0_and_A}. In the limit of vanishing $g$, it becomes $m^2_{\delta\phi}\approx\langle m_{3/2}\rangle^2\approx 27\mu^2/16$.

\begin{figure}
\centering
\begin{subfigure}{.47\textwidth}
  \centering
  \includegraphics[width=1\linewidth]{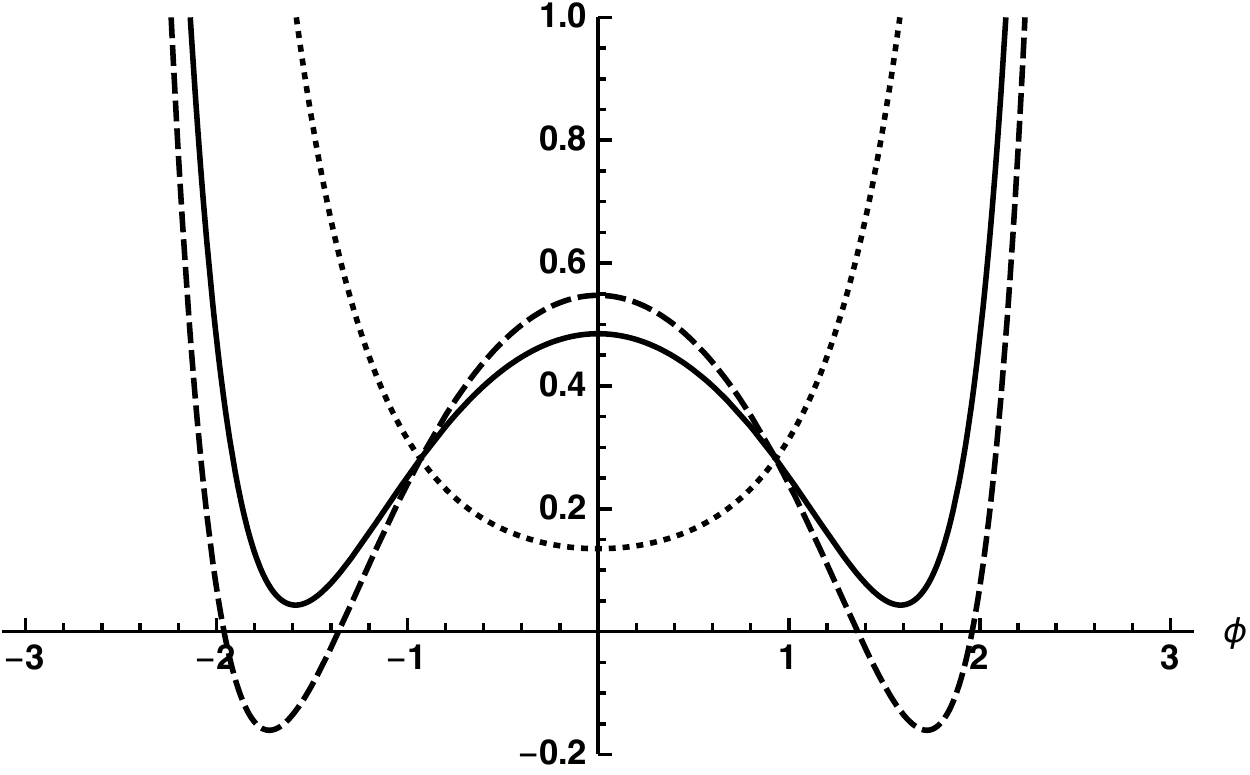}
  \caption{The case $\alpha=1$ and $g=0.5$. Solid line corresponds to $\mu=0.6$, dashed line to $\mu=0.65$, and dotted line to $\mu=0.1$.}
  \label{alpha1}
\end{subfigure}
\hspace{1em}
\begin{subfigure}{.47\textwidth}
  \centering
  \includegraphics[width=1\linewidth]{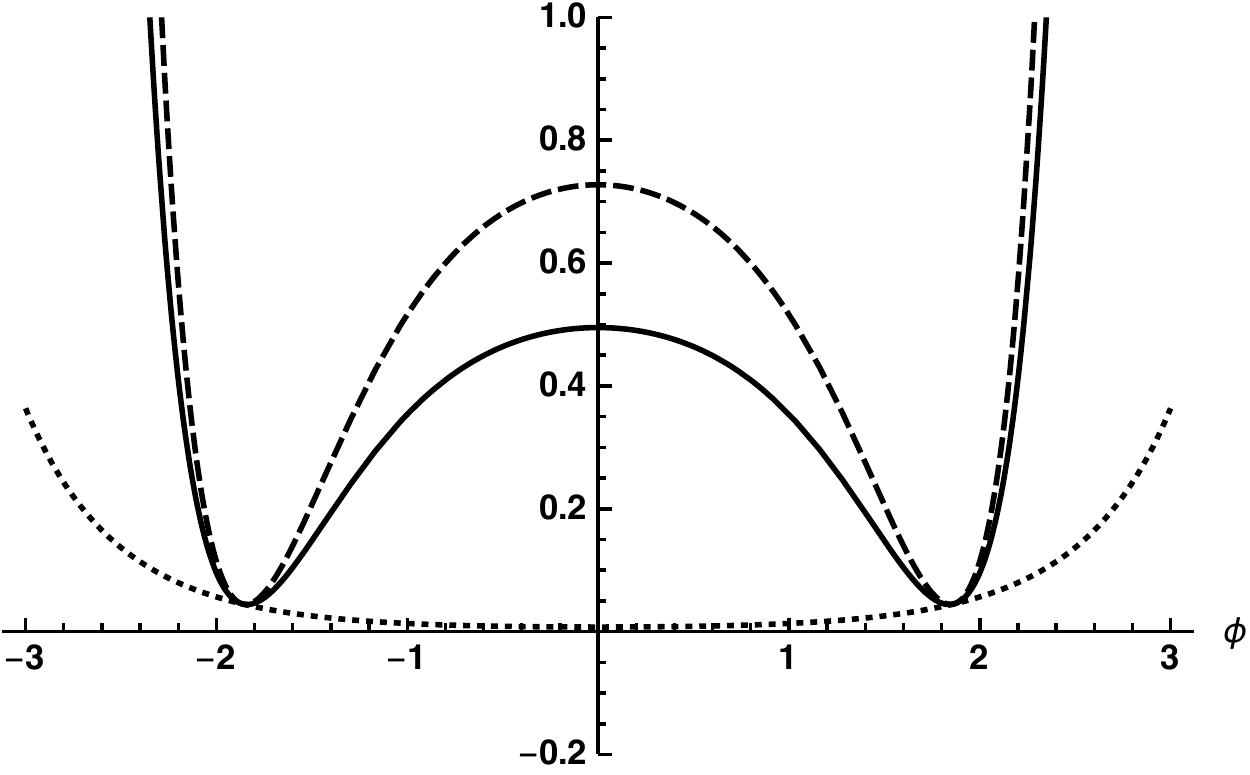}
  \caption{The case $\alpha=4$ and $g=0.1$. Solid line corresponds to $\mu=1.4$, dashed line to $\mu=1.7$, and dotted line to $\mu=0.1$.}
  \label{alpha4}
\end{subfigure}

\begin{subfigure}{.47\textwidth}
  \centering
  \includegraphics[width=1\linewidth]{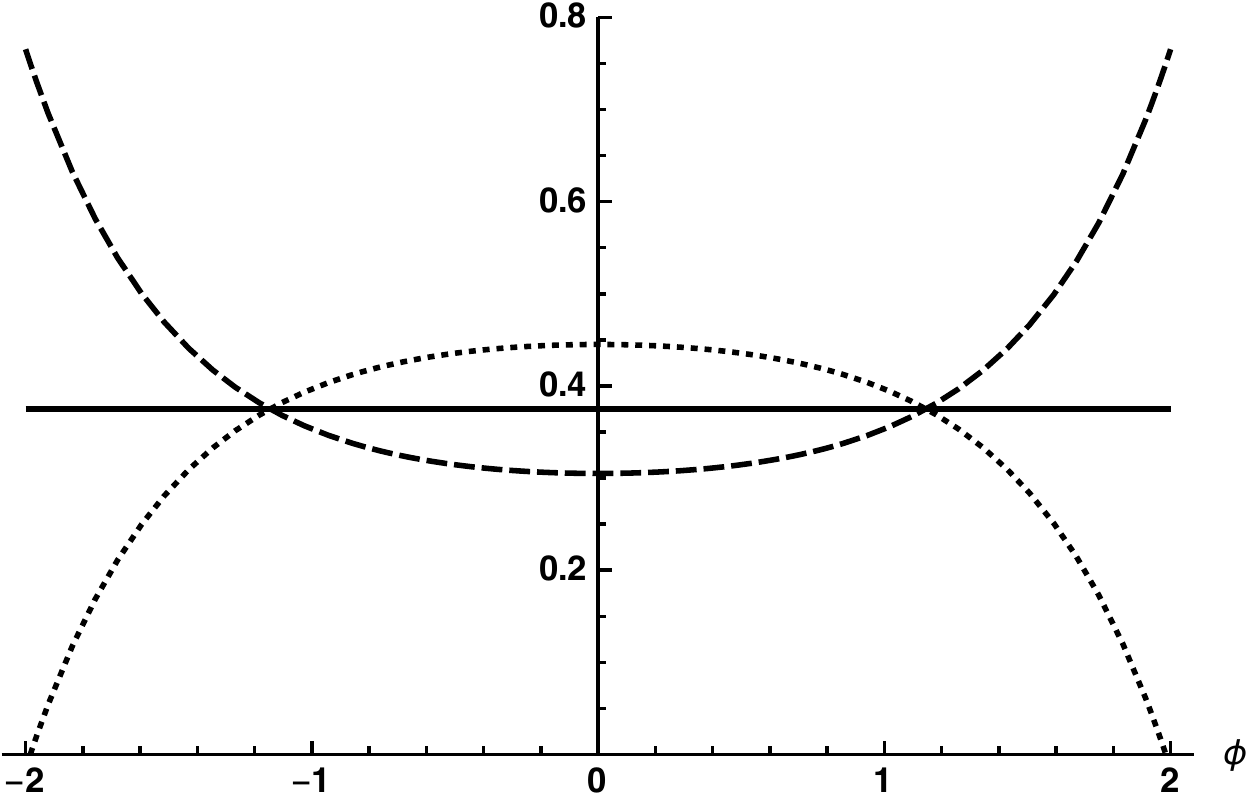}
  \caption{The case $\alpha=2$ and $g=0.5$. Solid line corresponds to $\mu=\sqrt{2}g\approx 0.707$ (no-scale choice), dashed line to $\mu=0.6$, and dotted line to $\mu=0.8$.}
  \label{alpha2}
\end{subfigure}
\hspace{1em}
\begin{subfigure}{.47\textwidth}
  \centering
  \includegraphics[width=1\linewidth]{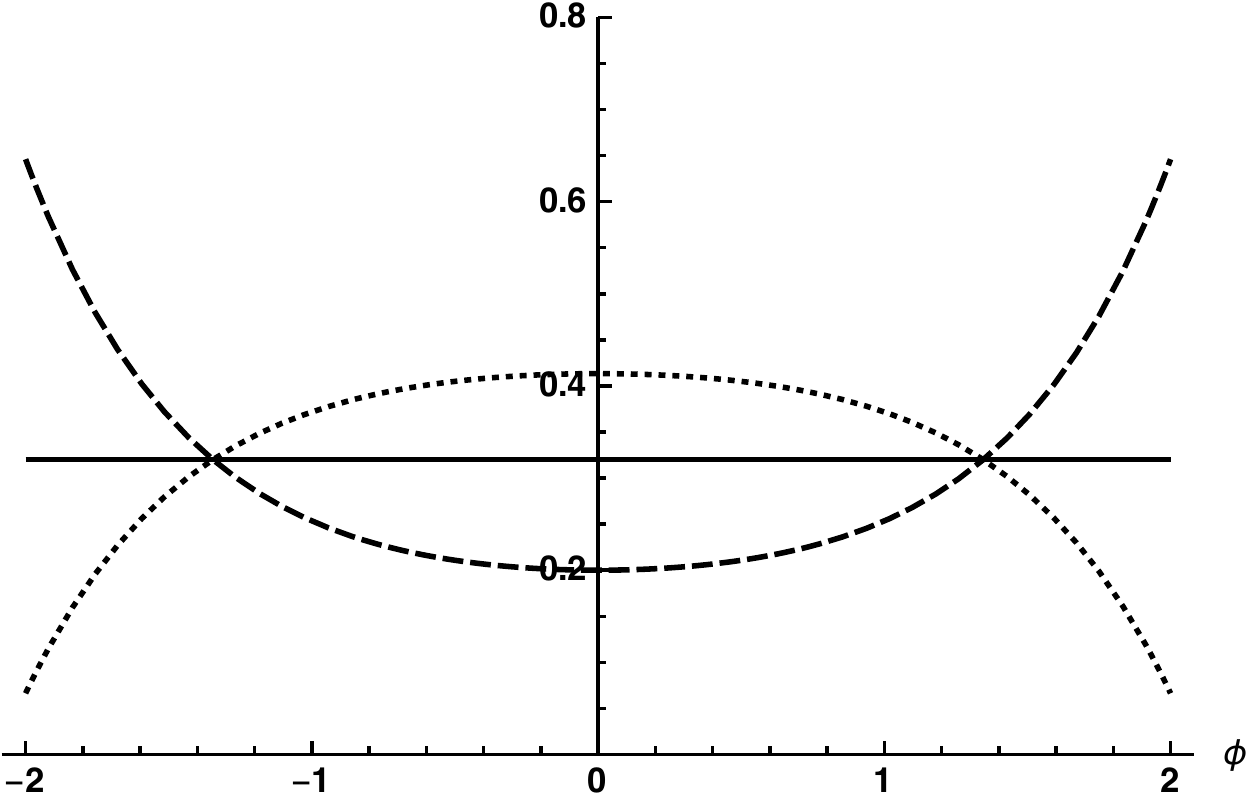}
  \caption{The case $\alpha=3$ and $g=0.4$. Solid line corresponds to $\mu=3g/\sqrt{2}\approx 0.849$ (no-scale choice), dashed line to $\mu=0.6$, and dotted line to $\mu=1$.}
  \label{alpha3}
\end{subfigure}
\captionsetup{width=1\linewidth}
\caption{Scalar potential $V(\phi)$, where $\phi$ is the canonical scalar, for $\alpha=1,2,3,4$ and different choices of the parameters $\mu$ and $g$.}
\label{Fig}
\end{figure}

For illustration purposes we provide the plots of the scalar potential for $\alpha=1,2,3,4$ in Figure \ref{Fig}.

\section{Generalizations}

\begin{figure}
\centering
\begin{subfigure}{.46\textwidth}
  \centering
  \includegraphics[width=1\linewidth]{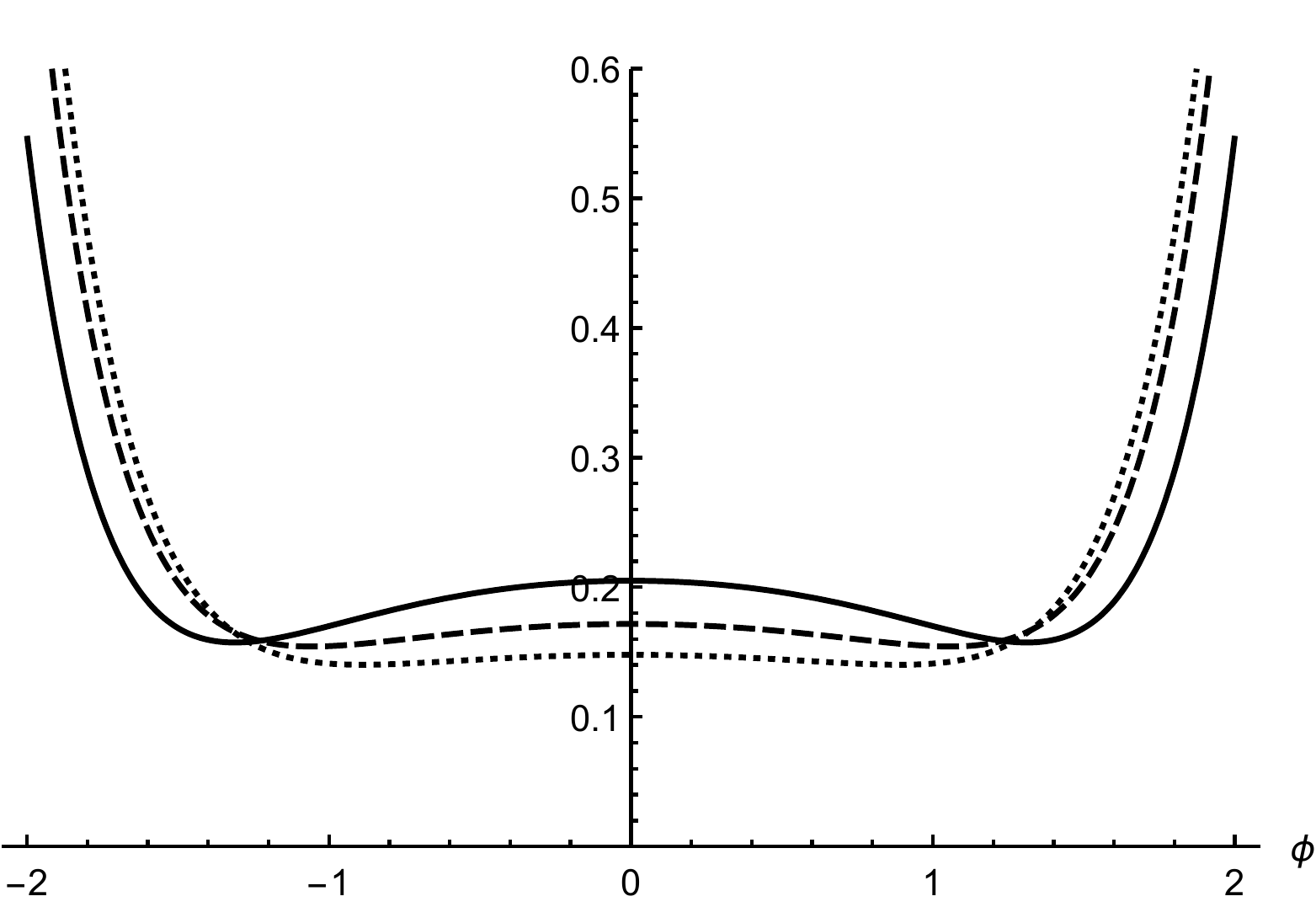}
  \caption{$\alpha=5$ (solid line), $\alpha=6$ (dashed line), and $\alpha=7$ (dotted line). The parameters values are $\mu=1$ and $g=0.1$ in all three cases.}
  \label{alpha567}
\end{subfigure}
\hspace{1em}
\begin{subfigure}{.49\textwidth}
  \centering
  \includegraphics[width=1\linewidth]{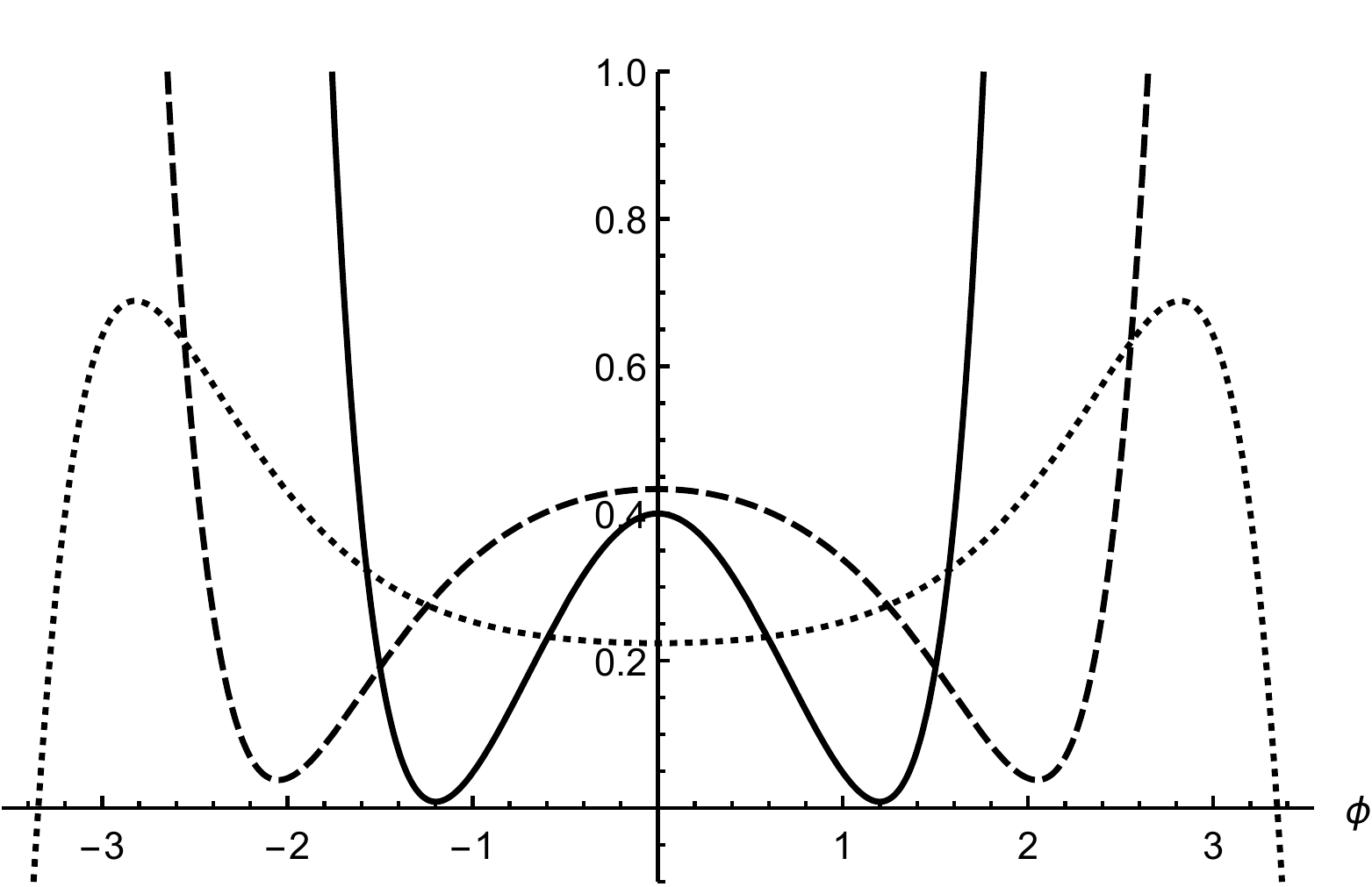}
  \caption{Solid line: $\alpha=1/2$, $\mu=0.4$, $g=0.4$. Dashed line: $\alpha=3/2$, $\mu=0.68$, $g=0.5$. Dotted line: $\alpha=5/2$, $\mu=0.6$, $g=0.4$.}
  \label{alphafrac}
\end{subfigure}
\captionsetup{width=1\linewidth}
\caption{Scalar potential for $\alpha=5,6,7$ (a) and $\alpha=1/2,3/2,5/2$ (b).}
\label{Fig2}
\end{figure}

Let us generalize $\alpha$, and recall the equation for critical points \eqref{zcrit},
\begin{equation}
    \alpha^2g^2(1-z^2)^\alpha+\mu^2(1-z^2)^2\left((\alpha-2)(\alpha-1)z^2-2\right)=0~.\label{zcrit2}
\end{equation}
It is convenient to introduce the notation
\begin{gather}
	1-z^2\equiv Y~,\nonumber\\
	(\alpha-1)(\alpha-2)-2\equiv B_1~,\\
	(\alpha-1)(\alpha-2)\equiv B_2~,\nonumber
\end{gather}
and rewrite Eq. \eqref{zcrit2} as
\begin{equation}
	\alpha^2g^2Y^\alpha+\mu^2 B_1 Y^2-\mu^2 B_2 Y^3=0~.\label{zcrit3}
\end{equation}
The no-scale structure can arise when (a) $B_1$ (or $B_2$) vanishes and (b) the remaining powers of $Y$ coincide, namely $\alpha=3$ (or $\alpha=2$). Then, since $Y$ cannot vanish (because $Y=1-z^2$ and $z=\tanh(\phi/\sqrt{2\alpha})$), Eq. \eqref{zcrit3} reduces to a relation between the parameters $\mu$ and $g$, that, if satisfied, leads to flatness of the potential. $B_1$ vanishes for $\alpha=0,3$, while $B_2$ vanishes for $\alpha=1,2$. Thus, for $\alpha=2,3$ the both conditions (a) and (b) are satisfied, and no-scale potential can be obtained. For other values of $\alpha$ flatness of the potential cannot be achieved (as long as $\mu,g\neq 0$) because all three powers of $Y$ in Eq. \eqref{zcrit3} are present and distinct. However, SUSY may still be broken by fixed VEVs of $z$ (or $Y$) as in the cases $\alpha=1,4$ that we studied. In Figure \ref{Fig2} we include plots of scalar potentials with three critical points, obtained for $\alpha=5,6,7$ (Figure \ref{alpha567}) and also fractional values $\alpha=1/2,3/2,5/2$ (Figure \ref{alphafrac}). As can be seen, certain parameter values of $\mu$ and $g$ allow for double-well potentials (with tunable minimum $V_0$) in all the above cases except $\alpha=5/2$ where the two $z\neq 0$ critical points become maxima rather than minima, and the potential is unbounded from below.

\begin{figure}
\centering
\includegraphics[width=.5\linewidth]{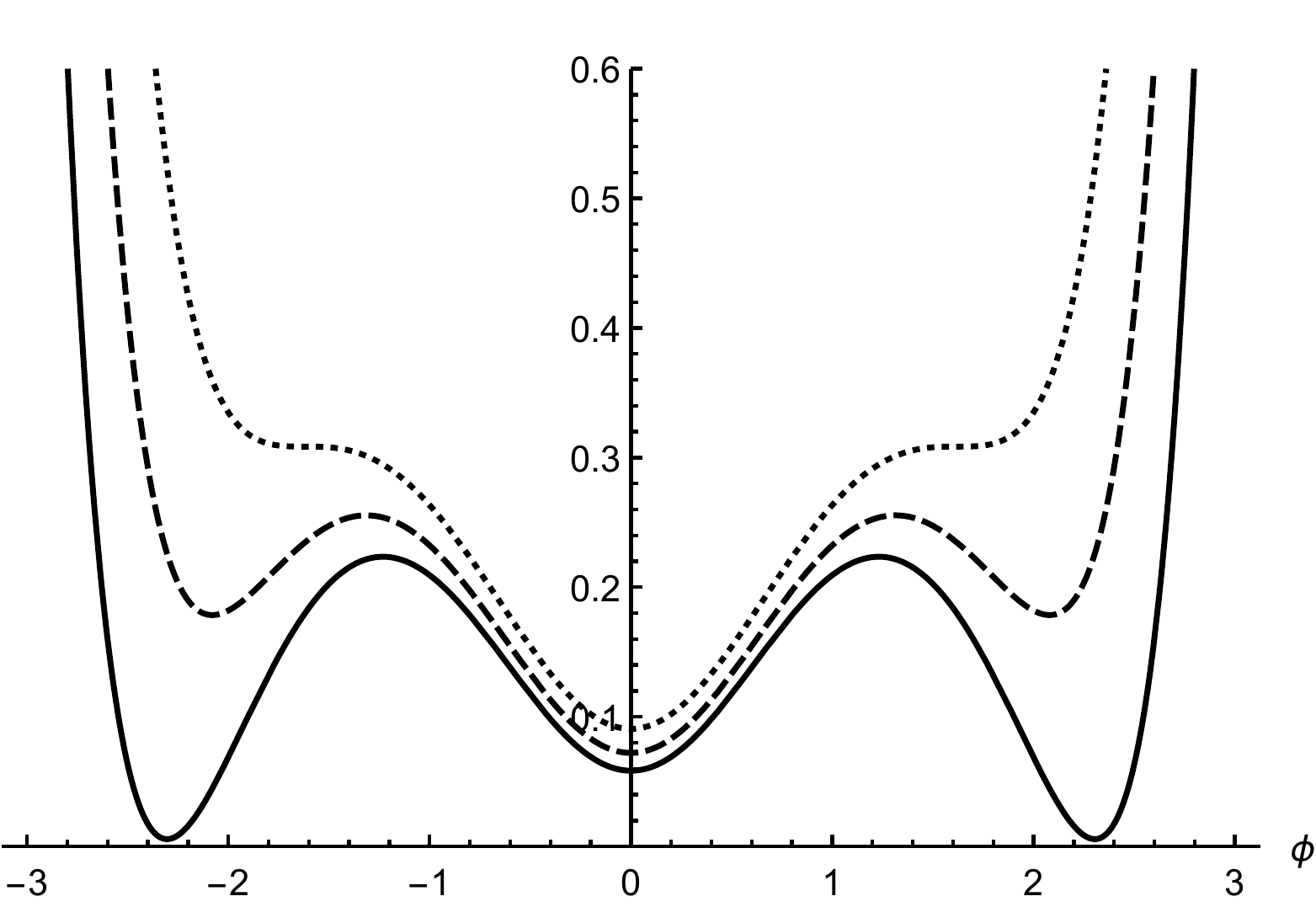}
\caption{Scalar potential for $\alpha=1$, $n=2$, and $\mu=0.35$. Solid line represents $g=0.171$, dashed line $g=0.19$, and dotted line $g=0.213$.}
\label{Fig3}
\end{figure}

As regards the generalization of $n$ in the superpotential \eqref{WZn}, it leads to the following equation for critical points,
\begin{multline}
\alpha^2g^2(1-z^2)^\alpha+\mu^2(1-z^2)^2z^{2n-4}[n(1-z^2)(n-1-z^2-nz^2)\\+\alpha z^2(2n-2-z^2-2nz^2)+\alpha^2z^4]=0~,\label{zcritn}
\end{multline}
that is a generalization of Eq. \eqref{zcrit2}. This introduces more diversity to the vacuum structure of the models. For example, taking $\alpha=1$ and $n=2$ we demonstrate in Figure \ref{Fig3} the case with five critical points (i.e. with Eq. \eqref{zcritn} having four real solutions with $0<z<1$). We fix $\mu=0.35$, and consider three values of $g$. When $g=0.171$ (solid line in Figure \ref{Fig3}) we have two maxima, one metastable minimum (false vacuum) at $z=\phi=0$ with preserved SUSY and $U(1)_R$, and two stable minima (true vacua) at $z\neq 0$ with broken SUSY and $U(1)_R$. In such scenario domain walls may form that divide the vacua with broken and unbroken SUSY and $U(1)_R$, depending on relative height of stable and metastable minima. The domain wall "bubbles" would be metastable and eventually decay~\footnote{This can leave stable domain walls that divide true vacua with $z=+|z_0|$ and $z=-|z_0|$.}, as the true vacuum with $z\neq 0$ is energetically favoured. For $g=0.19$ (dashed line in Figure \ref{Fig3}), on the other hand, the $z=0$ minimum becomes stable while $z\neq 0$ minima become metastable. In this case the decay of the domain walls would restore SUSY and R-symmetry. Finally, for $g=0.213$ (dotted line in Figure \ref{Fig3}) we have a single stable minimum at $z=0$, and two inflection points. When $g>0.213$ Eq. \eqref{zcritn} does not admit real solutions with $0<z<1$, so the $z\neq 0$ critical points disappear.

\section{Discussion and conclusion}

We constructed new models of spontaneous supersymmetry and R-symmetry breaking, based on $N=1$ four-dimensional supergravity coupled to a chiral multiplet with $SU(1,1)/U(1)$ (Poincar\'e plane) target space. The crucial part of our construction is gauged $U(1)_R$ symmetry that acts linearly on the Poincar\'e disk variable $Z$. This allows for SUSY breaking in de Sitter vacuum for appropriate parameter ranges.

More specifically, we considered the K\"ahler potential and superpotential
\begin{equation}
    K=-\alpha\log(1-Z\overbar{Z})~,~~~W=\mu Z~,\label{KW_setup}
\end{equation}
with integer values of $\alpha$ motivated by string theory constructions. We found that when $\alpha=1,4$, SUSY and R-symmetry are spontaneously broken provided that $2\mu^2>g^2$ (if $\alpha=1$) and $\mu^2>8g^2$ (if $\alpha=4$). In both cases positive cosmological constant can be generated. For $\alpha=2$ and $\alpha=3$ the situation is different -- for the specific choices $\mu^2=2g^2$ and $2\mu^2=9g^2$, respectively, we have flat potentials with positive tunable height. Consequently, the VEV of $Z$ is classically undetermined (to be fixed by perturbative corrections), and the SUSY breaking scale is arbitrary (with some restrictions), i.e. these two cases are examples of de Sitter no-scale supergravity. We also demonstrated that other values of $\alpha$ (including fractional ones) may lead to spontaneous SUSY and R-symmetry breaking as well, but the no-scale structure remains unique to $\alpha=2,3$. 

We discussed the generalization of $n$ in the superpotential $W=\mu Z^n$, and showed that it can generate potentials with more that two local minima, which can lead to some interesting implications such as formation of metastable domain wall bubbles that can decay into true vacua with broken or unbroken supersymmetry and R-symmetry, depending on the values of $\mu$ and $g$.

The tree-level spectrum of the models (after SUSY and R-symmetry breaking) consists of a massive vector, massive spin-1/2 field, and a massive real scalar (except for the no-scale cases where the potential is to be generated at one loop). The spin-1/2 field is a linear combination of the chiral fermion $\chi$ (superpartner of $Z$) and the gaugino $\lambda$, orthogonal to the goldstino. The $\chi$ and $\lambda$ have $U(1)_R$ charges $q(\chi)=q(\lambda)=1/2$, and therefore the pure model contains anomalies that must be cancelled after including the Supersymmetric Standard Model (SSM) and other possible fields. Also, the $U(1)_R$ gauge symmetry introduces a non-trivial task of assigning appropriate R-charges to the fields. For example, if the full superpotential is the sum $\mu Z+W_{\rm SSM}$, then the Standard Model R-charge assignments can be done along the lines of Ref. \cite{Chamseddine:1995gb}. Alternatively, $W_{\rm SSM}$ can be coupled to some power of $Z$ and thus carry different R-charge, or even be neutral.

We also checked whether or not viable single-field (hilltop) inflation can be realized with the models where $\alpha=1$ and $\alpha=4$ (with $n=1$). Unfortunately, it does not seem to be possible because the curvature of the potential around its maximum is too large. To be specific, for $\alpha=1$ the slow-roll parameter $\eta_*$ is 
\begin{equation}
    \eta_*\equiv\frac{V''(\phi_*)}{V(\phi_*)}\approx -1~,
\end{equation}
taken at the initial value of $\phi$ which we assume to be $\phi_*\approx 0$ (close to the maximum of the potential). Meanwhile the parameter
\begin{equation}
    \epsilon_*\equiv\frac{1}{2}\left(\frac{V'(\phi_*)}{V(\phi_*)}\right)^2~
\end{equation}
can be made small if the initial value of $\phi$ is close enough to zero. This means that the spectral tilt $n_s=1+2\eta_*-6\epsilon_*$ takes the value $n_s\approx -1$ that is incompatible with CMB data, $n_s\approx 0.965$ (see e.g. PLANCK 2018 results \cite{Akrami:2018odb}). On the other hand, the $\alpha=4$ case predicts smaller value of $\eta_*$, namely $\eta_*\approx -0.5$, but the tilt becomes $n_s\approx 0$ which is still unsatisfactory.~\footnote{For values $\alpha=5,6,7$ the scalar $\phi$ cannot be identified with the inflaton, because requiring $V_0\sim 10^{-120}$ would imply unacceptably small inflationary (Hubble) scale of similar order as $V_0$, while for $\alpha=1/2,3/2$ the problem of large $\eta_*$ remains.}

The situation is somewhat similar to the construction of Refs. \cite{Antoniadis:2017gjr,Antoniadis:2019dpm} where the K\"ahler potential is canonical (plus a quartic term), while the superpotential is linear due to the requirement of local R-symmetry. In this model viable hilltop inflation becomes possible only after including certain higher-order corrections to the K\"ahler potential. It is therefore of interest to continue the investigation of inflationary scenario in our models after including corrections/modifications to the K\"ahler potential, compatible with local R-symmetry.

\section*{Acknowledgements}

Y.A. was supported by the CUniverse research promotion project of Chulalongkorn University under the grant reference CUAASC, and the Ministry of Education and Science of the Republic of Kazakhstan under the grant reference BR05236322.

\bibliography{Bibliography.bib}{}

\providecommand{\href}[2]{#2}\begingroup\raggedright\begin{thebibliography}{10}

\bibitem{Riess:1998cb}
{\bfseries Supernova Search Team} Collaboration, A.~G. Riess {\em et~al.},
  ``{Observational evidence from supernovae for an accelerating universe and a
  cosmological constant},'' \href{http://dx.doi.org/10.1086/300499}{{\em
  Astron. J.} {\bfseries 116} (1998) 1009--1038},
\href{http://arxiv.org/abs/astro-ph/9805201}{{\ttfamily arXiv:astro-ph/9805201
  [astro-ph]}}.

\bibitem{Perlmutter:1997zf}
{\bfseries Supernova Cosmology Project} Collaboration, S.~Perlmutter {\em
  et~al.}, ``{Discovery of a supernova explosion at half the age of the
  Universe and its cosmological implications},''
  \href{http://dx.doi.org/10.1038/34124}{{\em Nature} {\bfseries 391} (1998)
  51--54},
\href{http://arxiv.org/abs/astro-ph/9712212}{{\ttfamily arXiv:astro-ph/9712212
  [astro-ph]}}.

\bibitem{Townsend:1977qa}
P.~K. Townsend, ``{Cosmological Constant in Supergravity},''
\href{http://dx.doi.org/10.1103/PhysRevD.15.2802}{{\em Phys. Rev.} {\bfseries
  D15} (1977) 2802--2804}.

\bibitem{Polonyi:1977pj}
J.~Polonyi, ``{Generalization of the Massive Scalar Multiplet Coupling to the
  Supergravity}.'' {Hungary Central Inst. Res. KFKI-77-93 (1977, REC. JUL
  1978), 5 p., unpublished}.

\bibitem{Linde:2016bcz}
A.~Linde, ``{On inflation, cosmological constant, and SUSY breaking},''
  \href{http://dx.doi.org/10.1088/1475-7516/2016/11/002}{{\em JCAP} {\bfseries
  1611} no.~11, (2016) 002},
\href{http://arxiv.org/abs/1608.00119}{{\ttfamily arXiv:1608.00119 [hep-th]}}.

\bibitem{Aldabergenov:2017bjt}
Y.~Aldabergenov and S.~V. Ketov, ``{Higgs mechanism and cosmological constant
  in $N=1$ supergravity with inflaton in a vector multiplet},''
  \href{http://dx.doi.org/10.1140/epjc/s10052-017-4807-8}{{\em Eur. Phys. J.}
  {\bfseries C77} no.~4, (2017) 233},
\href{http://arxiv.org/abs/1701.08240}{{\ttfamily arXiv:1701.08240 [hep-th]}}.

\bibitem{Wess:1992cp}
J.~Wess and J.~Bagger, {\em {Supersymmetry and supergravity}}.
\newblock Princeton University Press, Princeton, NJ, USA,
1992.
\newblock

\bibitem{Freedman:2012zz}
D.~Z. Freedman and A.~Van~Proeyen, {\em {Supergravity}}.
\newblock Cambridge University Press, Cambridge, UK,
2012.
\newblock

\bibitem{Duff:2010ss}
M.~J. Duff and S.~Ferrara, ``{Generalized mirror symmetry and trace
  anomalies},'' \href{http://dx.doi.org/10.1088/0264-9381/28/6/065005}{{\em
  Class. Quant. Grav.} {\bfseries 28} (2011) 065005},
\href{http://arxiv.org/abs/1009.4439}{{\ttfamily arXiv:1009.4439 [hep-th]}}.

\bibitem{Duff:2010vy}
M.~J. Duff and S.~Ferrara, ``{Four curious supergravities},''
  \href{http://dx.doi.org/10.1103/PhysRevD.83.046007}{{\em Phys. Rev.}
  {\bfseries D83} (2011) 046007},
\href{http://arxiv.org/abs/1010.3173}{{\ttfamily arXiv:1010.3173 [hep-th]}}.

\bibitem{Ferrara:2016fwe}
S.~Ferrara and R.~Kallosh, ``{Seven-disk manifold, $\alpha$-attractors, and $B$
  modes},'' \href{http://dx.doi.org/10.1103/PhysRevD.94.126015}{{\em Phys.
  Rev.} {\bfseries D94} no.~12, (2016) 126015},
\href{http://arxiv.org/abs/1610.04163}{{\ttfamily arXiv:1610.04163 [hep-th]}}.

\bibitem{Cremmer:1983bf}
E.~Cremmer, S.~Ferrara, C.~Kounnas, and D.~V. Nanopoulos, ``{Naturally
  Vanishing Cosmological Constant in N=1 Supergravity},''
\href{http://dx.doi.org/10.1016/0370-2693(83)90106-5}{{\em Phys. Lett.}
  {\bfseries 133B} (1983) 61}.

\bibitem{Ellis:1983sf}
J.~R. Ellis, A.~B. Lahanas, D.~V. Nanopoulos, and K.~Tamvakis, ``{No-Scale
  Supersymmetric Standard Model},''
\href{http://dx.doi.org/10.1016/0370-2693(84)91378-9}{{\em Phys. Lett.}
  {\bfseries 134B} (1984) 429}.

\bibitem{Ellis:1983ei}
J.~R. Ellis, C.~Kounnas, and D.~V. Nanopoulos, ``{Phenomenological SU(1,1)
  Supergravity},''
\href{http://dx.doi.org/10.1016/0550-3213(84)90054-3}{{\em Nucl. Phys.}
  {\bfseries B241} (1984) 406--428}.

\bibitem{Pallis:2018xmt}
C.~Pallis, ``{Gravity-mediated SUSY breaking, $R$ symmetry, and hyperbolic
  Kähler geometry},''
  \href{http://dx.doi.org/10.1103/PhysRevD.100.055013}{{\em Phys. Rev.}
  {\bfseries D100} no.~5, (2019) 055013},
\href{http://arxiv.org/abs/1812.10284}{{\ttfamily arXiv:1812.10284 [hep-ph]}}.

\bibitem{Chamseddine:1995gb}
A.~H. Chamseddine and H.~K. Dreiner, ``{Anomaly free gauged R symmetry in local
  supersymmetry},'' \href{http://dx.doi.org/10.1016/0550-3213(95)00583-8}{{\em
  Nucl. Phys.} {\bfseries B458} (1996) 65--89},
\href{http://arxiv.org/abs/hep-ph/9504337}{{\ttfamily arXiv:hep-ph/9504337
  [hep-ph]}}.

\bibitem{Akrami:2018odb}
{\bfseries Planck} Collaboration, Y.~Akrami {\em et~al.}, ``{Planck 2018
  results. X. Constraints on inflation},''
\href{http://arxiv.org/abs/1807.06211}{{\ttfamily arXiv:1807.06211
  [astro-ph.CO]}}.

\bibitem{Antoniadis:2017gjr}
I.~Antoniadis, A.~Chatrabhuti, H.~Isono, and R.~Knoops, ``{Inflation from
  Supersymmetry Breaking},''
  \href{http://dx.doi.org/10.1140/epjc/s10052-017-5302-y}{{\em Eur. Phys. J.}
  {\bfseries C77} no.~11, (2017) 724},
\href{http://arxiv.org/abs/1706.04133}{{\ttfamily arXiv:1706.04133 [hep-th]}}.

\bibitem{Antoniadis:2019dpm}
I.~Antoniadis, A.~Chatrabhuti, H.~Isono, and R.~Knoops, ``{A microscopic model
  for inflation from supersymmetry breaking},''
  \href{http://dx.doi.org/10.1140/epjc/s10052-019-7141-5}{{\em Eur. Phys. J.}
  {\bfseries C79} no.~7, (2019) 624},
\href{http://arxiv.org/abs/1905.00706}{{\ttfamily arXiv:1905.00706 [hep-th]}}.

\end{thebibliography}\endgroup
\bibliographystyle{utphys.bst}

\end{document}